\documentclass[fleqn,11pt,twoside]{article}
\usepackage{amsthm}
\usepackage[fleqn,intlimits]{amsmath}
\usepackage{graphicx}

\makeatletter
\newcommand{\copyrightnote}[2]{{\renewcommand{\thefootnote}{}
 \footnotetext{\small\it
\begin{flushleft}
Copyright \copyright \ #1 by  #2
\end{flushleft}}}}

\newcommand{\Name}[1]{\begin{flushleft}
                       \LARGE \bf #1
                       \end{flushleft}\vspace{-3mm}}

\newcommand{\Author}[1]{\begin{flushleft}
                       \it #1 \end{flushleft}}

\newcommand{\Address}[1]{\begin{flushleft}
                       \it #1 \end{flushleft}}

\newcommand{\FirstPageHead}[5]{
\begin{flushleft}
\raisebox{8mm}[0pt][0pt]
{\footnotesize \sf
\parbox{150mm}{ \qquad
 #1 #2 #3 
#4\hfill {\sc #5}}}\vspace{-13mm}
\end{flushleft}}

\newcommand{\evenhead}{Author \ name}
\newcommand{\oddhead}{Article \ name}

\renewcommand{\@evenhead}{
\hspace*{-3pt}\raisebox{-15pt}[\headheight][0pt]{\vbox{\hbox to \textwidth
{\thepage \hfil \evenhead}\vskip4pt \hrule}}}
\renewcommand{\@oddhead}{
\hspace*{-3pt}\raisebox{-15pt}[\headheight][0pt]{\vbox{\hbox to \textwidth
{\oddhead \hfil \thepage}\vskip4pt\hrule}}}
\renewcommand{\@evenfoot}{}
\renewcommand{\@oddfoot}{}

\long\def\@makecaption#1#2{%
  \vskip\abovecaptionskip
  \sbox\@tempboxa{\small \textbf{#1.}\ \ #2}%
  \ifdim \wd\@tempboxa >\hsize
    {\small \textbf{#1.}\ \ #2}\par
  \else
    \global \@minipagefalse
    \hb@xt@\hsize{\hfil\box\@tempboxa\hfil}%
  \fi
  \vskip\belowcaptionskip}

\newcommand{\JNMPnumberwithin}[3][\arabic]{%
  \@ifundefined{c@#2}{\@nocounterr{#2}}{%
    \@ifundefined{c@#3}{\@nocnterr{#3}}{%
      \@addtoreset{#2}{#3}%
      \@xp\xdef\csname the#2\endcsname{%
        \@xp\@nx\csname the#3\endcsname .\@nx#1{#2}}}}%
}

\newcommand{\resetfootnoterule} {
  \renewcommand\footnoterule{%
  \kern-3\p@
  \hrule\@width.4\columnwidth
  \kern2.6\p@}
}

\renewcommand{\footnoterule}{}

\newcommand{\be}{\begin{equation}}
\newcommand{\ee}{\end{equation}}
\newcommand{\ba}{\hspace*{-5pt}\begin{array}}
\newcommand{\ea}{\end{array}}
\newcommand{\p}{\partial}

\makeatother

\numberwithin{equation}{section}
\theoremstyle{definition}

\renewcommand{\ba}{\begin{array}}
\renewcommand{\ea}{\end{array}}
\newcommand{\beg}{\begin{eqnarray}}
\newcommand{\eeq}{\end{eqnarray}}
\newcommand{\bg}{\begin{eqnarray*}}

\newcommand{\ed}{\end{eqnarray*}}
\newcommand{\n}{\newline\hfill}

\renewcommand{\p}{\partial} 
 
\newcommand{\notlhd}{\lhd\kern-.8em{/}\ } 
\newcommand{\notexist}{\ \exists\kern-.5em{\raise.1em\hbox{/}}\ }

\newcommand{\pde}[2]{\frac{\p #1}{\p #2}}

\newcommand{\inp}{{\mbox{\vbox{\hrule width0ex\hbox{\vrule
 height0ex\kern3.8pt
\vbox{\kern2.5pt}\kern3.8pt \vrule height1.6ex}
\hrule width1.6ex}}}}

\begin{document}

\renewcommand{\evenhead}{N Petersson, N Euler and M Euler}
\renewcommand{\oddhead}{Integrable Third-Order Evolution Equations}


\thispagestyle{empty}

\begin{flushleft}
\footnotesize \sf
\end{flushleft}

\FirstPageHead{\ }{\ }{\ }
{ }{{
{ }}}
\copyrightnote{2003}{N Petersson, N Euler and M Euler}

\Name{Recursion Operators for a
Class of Integrable Third-Order Evolution Equations}

\label{euler-firstpage}

\Author{Niclas PETERSSON, Norbert EULER$^*$ and Marianna EULER}

\Address{Department of Mathematics,  Lule\aa\ University of Technology, \\
SE-971 87 Lule\aa, Sweden\\\quad \\
$^*$ Corresponding author's e-mail: norbert@sm.luth.se}

\strut\hfill

\begin{abstract}
\noindent
We consider $u_t=u^{\alpha} u_{xxx}+n(u)u_xu_{xx}+m(u)u_x^3+ r(u)u_{xx}
+p(u)u_x^2 + q(u)u_x+s(u)$ with $\alpha=0$ and $\alpha=3$,
for those functional forms of $m,\ n,\ p,\ q,\ r,\ s$
for which the equation is
integrable in the sense of an infinite number of Lie-B\"acklund
symmetries.
Local $x$- and $t$-independent recursion operators that generate
these infinite sets of symmetries
are obtained for the equations. A combination of
potential forms, hodograph transformations and $x$-generalised hodograph
transformations are applied to the obtained equations.

\end{abstract}



\section{Introduction}

A partial differential equation that admits a recursion operator
which generates an infinite
number of local $x$- and $t$-independent Lie-B\"acklund symmetries
is called integrable \cite{fokas80, fokas87}.

In the present paper we classify the $(1+1)$-dimensional
semilinear evolution equation of the form
\begin{gather}
\label{gen_eq}
u_t=u_{xxx}+n(u)u_xu_{xx}+m(u)u_x^3+ r(u)u_{xx} +p(u)u_x^2+ q(u)u_x+s(u),
\end{gather}
as well as the quasilinear evolution equation of the form
\begin{gather}
\label{gen_quasi}
u_t=u^3u_{xxx}+n(u)u_xu_{xx}+m(u)u_x^3+ r(u)u_{xx} +p(u)u_x^2+ q(u)u_x+s(u),
\end{gather}
where the $C^3$-functions $m,\ n,\ p,\ q,\ r,\ s$
are constrained by the
existence of a local recursion operator of the form
\begin{gather}
\label{gen_R}
R[u]=GD_x^2+QD_x+H+\sum_jI_jD_x^{-1}J_j.
\end{gather}
Here the functions $G,\ Q,\ H,\ I_j,\ J_j$ depend on
$u$ and $x$-derivatives of $u$ and are determined for
(\ref{gen_eq}) and (\ref{gen_quasi})
by the commutator relation
\begin{gather}
\label{commutator}
\left[ \vphantom{a^{x^2}}
L[u],\ R[u]\right]\varphi =\left(\vphantom{\frac{2}{3}}D_tR[u]\right)\varphi,
\end{gather}
where $L[u]$ is the linear operator
\begin{gather}
L[u]=\pde{F}{u}+\pde{F}{u_x}D_x+\pde{F}{u_{xx}}D_x^2+\cdots +
\pde{F}{u_{x^n}}D_x^n,
\end{gather}
$F$ is the right-hand-side of (\ref{gen_eq}) or (\ref{gen_quasi})
and $D_tR[u]$ calculates the explicit derivative with respect to $t$.
For more details on recursion operators
we refer to \cite{fokas80, fokas87}.

In two recent papers \cite{eul_eul, eul_eul_pet} we classified
$(1+1)$-dimensional autonomous evolution
equations which
can be transformed to linear evolution equations
under the $x$-generalised hodograph transformation
\begin{gather}
\label{n_hodo}
_n\mbox{\bf H}:\left\{\ba{l}
\displaystyle{dX(x,t)
=f_1(x,u)dx+f_2(x,u,u_{x}, u_{xx},\ldots
, u_{x^{n-1}})dt}
\\[3mm]
dT(x,t)=dt \\[3mm]
U(X,T)=g(x)
\ea\right.
\end{gather}
with $n=2,3,\ldots$, where
\begin{gather}
\label{poincare}
u_{t}\pde{f_1 }{u}=
\pde{f_2}{x}+u_{x}\pde{f_2 }{u}+
u_{xx}\pde{f_2 }{u_{x}}
+\cdots +u_{x^{n}}\pde{f_2}{u_{x^{n-1}}}.
\end{gather}
We call these equations linearisable.
For second-order equations the $x$-generalised hodograph
transformation $_2\mbox{\bf H}$
leads to eight linearisable equations \cite{eul_eul}. We list here those
linearisable equations which admit 
$x$- and $t$-independent local recursion operators \cite{eul_eul_pet}:
\begin{gather}
\label{Eq2_1}
u_t=u_{xx}+\left(\frac{k''}{k'}+\alpha k'\right)u_x^2+\beta u_x
+\frac{\gamma}{k'}\\
\label{R2_1}
R[u]=D_x+\left(\frac{k''}{k'}+\alpha k'\right)u_x
\end{gather}
\begin{gather}
\label{Eq2_2}
u_t=u_{xx}+\frac{k''}{k'}u_x^2+\alpha ku_x+\beta u_x\\
\label{R2_2}
R[u]=D_x+\frac{k''}{k'}u_x+\frac{\alpha}{2}k+\frac{\alpha}{2}u_xD_x^{-1}k'
\end{gather}
\begin{gather}
\label{Eq2_3}
u_t=k^2u_{xx}+k^2\frac{k''}{k'}u_x^2+\alpha k^2u_x+\beta u_x\\
\label{R2_3}
R[u]=kD_x+k\frac{k''}{k'}u_x+\alpha k+
\left(k^2u_{xx}+k^2\frac{k''}{k'}u_x^2+\alpha k^2u_x\right)D_x^{-1}
\frac{k'}{k^2}.
\end{gather}
Here $\alpha,\ \beta,\ \gamma$ are arbitrary constants and
$k$ is a nonconstant ${\cal C}^3$ function of $u$.
In the present paper we exclude
these linearisable cases already listed in \cite{eul_eul_pet}.

\strut\hfill

The article is organized as follows:
In Section 2 and Section 3 we classify the integrable
equations of the form (\ref{gen_eq}) and (\ref{gen_quasi})
with respect to their local recursion operators.
The results are presented in Proposition 2.1 and Proposition 3.1.
In all cases we give the next equations in the integrable hierarchies,
which are all fifth-order equations.
Several examples are given and all equations are presented
in a more general form under the substitution $u\to k(u)$.
In Section 4 we consider sequences of coordinate transformations
for the integrable equations provided by Propositions 2.1 and
3.1, namely combinations of
potential forms, hodograph transformations and $x$-generalised
hodograph transformations. As mentioned above, we are interested
in local recursion oparators. However, we make one exception in
Subsection 4.4, where we derive a nonlocal recursion operator
for eq. (\ref{CDIS}),
which is a linearisable third-order equation.
This provides an interesting example of nonlocal recursion operators
and its corresponding nonlocal symmetries and sets the stage for future
investigations. In Section 5 we make some conclusions and
list all equations for which we established
recursion operators.

\section{Recursion operators for the semilinear equation (\ref{gen_eq})}

The recursion operator Ansatz (\ref{gen_R}) for (\ref{gen_eq})
leads to

\strut\hfill

\noindent
{\bf Proposition 2.1:} {\it Evolution equations of the form
(\ref{gen_eq}), {\it viz.}
\begin{gather*}
u_t=u_{xxx}+n(u)u_xu_{xx}+m(u)u_x^3+ r(u)u_{xx} +p(u)u_x^2+ q(u)u_x+s(u)
\end{gather*}
which admit recursion operators of the form (\ref{gen_R}), {\it viz.}
\begin{gather*}
R[u]=GD_x^2+QD_x+H+\sum_jI_jD_x^{-1}J_j
\end{gather*}
and which are not linearisable by a $x$-generalised hodograph transformation, 
are exhausted by the following two cases:\n
\begin{enumerate} 
\item[{\bf I)}]
The equation
\begin{gather}
\boxed{
\label{gen_eq3}
u_t=u_{xxx}+\lambda_1u_x^3+\lambda_2 u_x^2+\lambda_3u_x+\lambda_4
}
\end{gather}
with $\lambda_1,\ \lambda_2,\ \lambda_3,\ \lambda_4$ arbitrary
constants,
admits the recursion operator
\begin{gather}
\label{R_c}
R[u]=D_x^2+2\lambda_1 u_x^2+\frac{4\lambda_2}{3}u_x
-2\lambda_1u_xD_x^{-1}u_{xx}
-\frac{2\lambda_2}{3}D_x^{-1}u_{xx}.
\end{gather}
\item[{\bf II)}] The equation
\begin{gather}
\label{gen_eq1}
\boxed{
u_t=u_{xxx}+n(u)u_xu_{xx}+m(u)u_x^3+q(u)u_x
}
\end{gather}
admits the recursion operator (\ref{gen_R}), where
\begin{gather*}
Q=\frac{2}{3}nu_x,\qquad G=1\\
H=\frac{1}{3}nu_{xx}-\frac{1}{9}\left(3n'+n^2-18m\right)u_x^2+\frac{2}{3}q\\
I=u_x,\\
J=\frac{2}{9}\left(3n'+n^2-9m\right)u_{xx}+\frac{1}{9}\left(
3n''+2nn'-9m'\right)u_x^2+\frac{1}{3}q'
\end{gather*}
and $m,\ n,\ p,\ q,\ s$ 
satisfy the following conditions:
\begin{gather}
\label{gen_eq1_1}
q'''-nq''-3n'q'-\frac{2}{3}n^2q'+8mq'=0\\
\label{gen_eq1_2}
9m'-3n''-6nm+\frac{2}{3}n^3=0.
\end{gather} 
\end{enumerate}
}

\noindent
To prove Proposition 2.1 one needs to
verify commutator realtion
(\ref{commutator}) for eq. (\ref{gen_eq}) and the recursion operator
Ansatz (\ref{gen_R}). This is done by direct calculations, so we do
not show the details here.

\strut\hfill

In order to obtain the next equations in the integrable hierarchies of
the equations in Proposition 2.1 we apply the recursion operators on
the $t$-translation symmetries, i.e. $R[u]u_t$.
For eq. (\ref{gen_eq3}) the next equation in the
hierarchy is the fifth-order equation
\begin{gather}
u_t=u_{xxxxx}+\lambda_2\left(
\frac{10}{3}u_xu_{xxx}+\frac{5}{3}u_{xx}^2+\frac{10\lambda_2}{9}
u_x^3\right)+\lambda_3\left(u_{xxx}+\lambda_1
u_x^3+\lambda_2u_x^2\right)
\notag\\
\qquad
\label{gen_eq3f}
+\frac{2\lambda_4}{3}u_x
+\lambda_1\left(5u_x^2u_{xxx}+5u_xu_{xx}^2
+\frac{3\lambda_1}{2}u_x^5+\frac{5\lambda_2}{2}u_x^4\right).
\end{gather}

We note that eq. (\ref{gen_eq3}) of Proposition 2.1
contains the well known
potential KdV equation (with $\lambda_1=\lambda_4=0$)
as well as the potential modified KdV
equation (with $\lambda_2=\lambda_4=0$).
Equation (\ref{gen_eq3}) and its recursion operator has also been
reported in \cite{fokas80}.

\strut\hfill

\noindent
By the substitution
\begin{gather*}
u\to k(u),\qquad k\in {\cal C}^3
\end{gather*}
eq. (\ref{gen_eq3}) takes the form
\begin{gather}
\label{gen_eq3k}
u_t=u_{xxx}+3\frac{k''}{k'}u_xu_{xx}
+\left(\frac{k'''}{k'}+\lambda_1(k')^2\right)u_x^3+
\lambda_2k'u_x^2+\lambda_3 u_x
+\frac{\lambda_4}{k'}
\end{gather}
and admits the recursion operator
\begin{gather}
R[u]=D_x^2+2\frac{k''}{k'}u_xD_x+\frac{k''}{k'}u_{xx}
+\left(\frac{k'''}{k'}+2\lambda_1(k')^2\right)u_x^2+\frac{4\lambda_2}{3}
k'u_x\notag\\
\label{gen_eq3R}
\qquad
-2\lambda_1u_xD_x^{-1}\left[(k')^2u_{xx}+k'k''u_x^2\right]
-\frac{2}{3}\frac{\lambda_2}{k'}D_x^{-1}\left[(k')^2u_{xx}+k'k''u_x^2\right].
\end{gather}

We now give some examples of equation (\ref{gen_eq1})
and its recursion operators of Case II in Proposition 2.1

\strut\hfill

\noindent
{\bf Example 2.1:} Let
\begin{gather*}
m(u)= -\frac{1}{54}u
\end{gather*}
and make use of the substitution $u(x,t)=3\tilde u(x,t)$. Then system
(\ref{gen_eq1_1}) -- (\ref{gen_eq1_2}) reduces to
\begin{gather}
\label{P_II}
n''=2n^3+n\tilde u-\frac{1}{2}\\
\label{q_3rd}
q'''-3nq'' -9\left(n'+\frac{2}{3}n^2+\frac{4}{9}\tilde u\right)q'=0.
\end{gather}
Here $'=d/d\tilde u$, etc.
We note that (\ref{P_II}) is the well known
second Painlev\'e transcendent.
Hence, system (\ref{P_II}) -- (\ref{q_3rd})
is solvable in general.
By Proposition 2.1 the equation
\begin{gather}
\label{CaseI_a}
\boxed{
\tilde u_t=\tilde u_{xxx}+3n(\tilde u)\tilde u_x\tilde
u_{xx}-\frac{1}{2}\tilde u\tilde u_x^3+q(\tilde u)\tilde u_x
}
\end{gather}
where $n$ and $q$ satisfy (\ref{P_II}) -- (\ref{q_3rd}),
admits the recursion operator
\begin{gather}
\label{CaseIa_R}
R[\tilde u]=D^2_x+2n\tilde u_xD_x+n\tilde u_{xx}
-(n'+n^2+\tilde u)\tilde u_x^2+\frac{2}{3}q\notag\\[0.3cm] 
\qquad+\tilde u_xD_x^{-1}\left[2\left(n'+n^2
+\frac{1}{2}\tilde u\right)\tilde u_{xx}
+\left(n''+2n'n+\frac{1}{2}\right)\tilde u_x^2+\frac{1}{3}q'\right].
\end{gather}
The next equation in the hierarchy is
\begin{gather}
\label{ex1_5}
\tilde u_t=\tilde u_{xxxxx}+5n\tilde u_x\tilde u_{xxxx}+10n\tilde
u_{xx}\tilde u_{xxx}+5\left(n'+n^2-\frac{1}{2}\tilde u\right)
\tilde u_x^2\tilde  u_{xxx}\notag\\
\qquad +10\left(n'+n^2-\frac{1}{4}\tilde u\right)\tilde u_x\tilde u_{xx}^2
+5\left(nn'+n^3-\frac{1}{2}\tilde un-1\right)\tilde u_x^3\tilde u_{xx}\notag\\
\qquad
+\frac{1}{2}\left(-(n')^2+n^4+\tilde  un^2-2n+\frac{3}{4}\tilde
u^2\right)\tilde u_x^5
+\frac{5}{3}q\tilde u_{xxx}\notag\\
\qquad
+5\left(\frac{2}{3}q'+nq\right)\tilde u_x\tilde u_{xx}
+\left(\frac{5}{6}q''+\frac{5}{2}nq'-\frac{1}{2}\tilde uq\right)\tilde
u_x^3
+\frac{5}{6}q^2\tilde u_x,
\end{gather}
where the conditions on $n$ and $q$
satisfy
(\ref{P_II}) -- (\ref{q_3rd}).

\strut\hfill

\noindent
{\bf Example 2.2:} Let
\begin{gather*}
m(u)=0,\qquad n(u)=3u^{-1}\\[0.3cm]
q(u)=\lambda_1u^4+\lambda_2 u^2+\lambda_3,\qquad
\lambda_1,\ \lambda_2,\ \lambda_3\in\Re,
\end{gather*}
which satisfies system (\ref{gen_eq1_1}) -- (\ref{gen_eq1_2}). Then
(\ref{gen_eq1}) takes the form
\begin{gather}
\label{CaseI_b}
\boxed{
u_t=u_{xxx}+3u^{-1}u_xu_{xx}+ \lambda_1u^4u_x+\lambda_2 u^2u_x+\lambda_3u_x 
}
\end{gather}
and by Proposition 2.1, eq. (\ref{CaseI_b}) admits the recursion operator
\begin{gather}
\label{CaseIb_R}
R[u]=D_x^2+2u^{-1}u_xD_x+u^{-1}u_{xx}+\frac{2\lambda_1}{3}u^4+
\frac{2\lambda_2}{3}u^2\notag\\
\qquad
+\frac{2}{3}u_xD_x^{-1}\left[2\lambda_1u^3+\lambda_2u\right].
\end{gather}
The next equation in the hierarchy is
\begin{gather}
u_t=u_{xxxxx}+5u^{-1}u_xu_{xxxx}+10u^{-1}u_{xx}u_{xxx}
+\frac{5\lambda_1}{3}\left(u^4u_{xxx}+11u^3u_xu_{xx}+12u^2u_x^3\right)
\notag\\[0.3cm]
\qquad +\frac{5\lambda_2}{3}\left(u^2u_{xxx}+7uu_xu_{xx}+4u_x^3\right)
+\lambda_3\left(u_{xxx}+3u^{-1}u_xu_{xx}+\lambda_1u^4u_x
+\lambda_2u^2u_x\right)\notag\\[0.3cm]
\label{ex1b_5}
\qquad
+\frac{5\lambda_1^2}{6}u^8u_x+\frac{5\lambda_1\lambda_2}{3}u^6u_x
+\frac{5\lambda_2^2}{6}u^4u_x.
\end{gather}
By the substitution
\begin{gather*}
u\to k(u),\qquad k\in {\cal C}^3,
\end{gather*}
eq. (\ref{CaseI_b}) takes the form
\begin{gather}
\label{CaseIb_k}
\boxed{
u_t=u_{xxx}
+3\left(\frac{k''}{k'}+\frac{k'}{k}\right)u_xu_{xx}
+\left(\frac{k'''}{k'}+3\frac{k''}{k}\right)u_x^3
+\lambda_1k^4u_x+\lambda_2k^2u_x+\lambda_3u_x
}
\end{gather}
so that system (\ref{gen_eq1_1}) -- (\ref{gen_eq1_2}) admits the
solution
\begin{gather*}
n(u)=3\left(\frac{k''}{k'}+\frac{k'}{k}\right),\qquad
m(u)=\frac{k'''}{k'}+3\frac{k''}{k}\\[0.3cm]
q(u)=\lambda_1 k^4+\lambda_2k^2+\lambda_3.
\end{gather*}
By Proposition 2.1, eq. (\ref{CaseIb_k}) admits the recursion operator
\begin{gather}
R[u]=D_x^2+2\left(\frac{k''}{k'}+\frac{k'}{k}\right)u_xD_x
+\left(\frac{k''}{k'}+\frac{k'}{k}\right)u_{xx}
+\left(\frac{k'''}{k'}+3\frac{k''}{k}\right)u_{x}^2\notag\\[0.3cm]
\label{CaseIb_Rk}
\qquad +\frac{2}{3}\left(\lambda_1k^4+\lambda_2 k^2\right)
+\frac{2}{3}u_xD_x^{-1}\left[2\lambda_1k^3k'+\lambda_2kk'\right].
\end{gather}

\strut\hfill

\noindent
{\bf Example 2.3:} Let
\begin{gather*}
m(u)=\frac{3}{2}u^{-2},\qquad n(u)=-3u^{-1}\\[0.3cm]
q(u)=\lambda_1u^2+\lambda_2 u^{-2}+\lambda_3,\qquad
\lambda_1,\ \lambda_2,\ \lambda_3\in\Re,
\end{gather*}
which satisfies system (\ref{gen_eq1_1}) -- (\ref{gen_eq1_2}). Then
(\ref{gen_eq1}) takes the form
\begin{gather}
\label{CaseI_bN}
\boxed{
u_t=u_{xxx}-3u^{-1}u_xu_{xx}+\frac{3}{2}u^{-2}u_x^3
+\lambda_1u^2u_x+\lambda_2 u^{-2}u_x+\lambda_3u_x 
}
\end{gather}
and by Proposition 2.1, eq. (\ref{CaseI_bN}) admits the recursion operator
\begin{gather}
\label{CaseIb_RN}
R[u]=D_x^2-2u^{-1}u_xD_x-u^{-1}u_{xx}+u^{-2}u_x^2
+\frac{2\lambda_1}{3}u^2+
\frac{2\lambda_2}{3}u^{-2}\notag\\[0.3cm]
\qquad
+u_xD_x^{-1}\left[u^{-2}u_{xx}-u^{-3}u_x^2+
\frac{2\lambda_1}{3}u-\frac{2\lambda_2}{3}u^{-3}\right].
\end{gather}
The next equation in the hierarchy is
\begin{gather}
u_t=u_{xxxxx}-5u^{-1}u_xu_{xxxx}-10u^{-1}u_{xx}u_{xxx}
+\frac{35}{2}u^{-2}u_x^2u_{xxx}+\frac{55}{2}u^{-2}u_xu_{xx}^2\notag\\[0.3cm]
\qquad
-\frac{95}{2}u^{-3}u_x^3u_{xx}+\frac{135}{8}u^{-4}u_x^5
+\frac{5\lambda_1}{3}\left(u^2u_{xxx}+uu_xu_{xx}-\frac{1}{2}u_x^3\right)
+\frac{5\lambda_2^2}{6}u^{-4}u_x
\notag\\[0.3cm]
\qquad
+\frac{5\lambda_2}{3}\left(u^{-2}u_{xxx}-7u^{-3}u_xu_{xx}+
\frac{15}{2}u^{-4}u_x^3\right)
+\frac{5\lambda_1^2}{6}u^4u_x+\frac{4\lambda_1\lambda_2}{3}u_x
\notag\\[0.3cm]
\qquad
+\lambda_3\left(u_{xxx}-3u^{-1}u_xu_{xx}+\frac{3}{2}u^{-2}u_x^3
+\lambda_1u^2u_x
+\lambda_2u^{-2}u_x\right).
\label{ex1b_5N}
\end{gather}
By the substitution
\begin{gather*}
u\to k(u),\qquad k\in {\cal C}^3,
\end{gather*}
eq. (\ref{CaseI_bN}) takes the form
\beg
\boxed{
\label{CaseIb_kN}
\ba{l}
\displaystyle{
u_t=u_{xxx}
+3\left(\frac{k''}{k'}-\frac{k'}{k}\right)u_xu_{xx}
+\left(\frac{k'''}{k'}-3\frac{k''}{k}
+\frac{3}{2}\frac{(k')^2}{k^2}\right)u_x^3}\\
[0.5cm]
\displaystyle{
\qquad +\lambda_1k^2u_x+\frac{\lambda_2}{k^2}u_x+\lambda_3u_x}
\ea
}
\eeq
so that system (\ref{gen_eq1_1}) -- (\ref{gen_eq1_2}) admits the
solution
\begin{gather*}
n(u)=3\left(\frac{k''}{k'}-\frac{k'}{k}\right),\qquad
m(u)=\frac{k'''}{k'}-3\frac{k''}{k}+\frac{3}{2}\frac{(k')^2}{k^2} \\
q(u)=\lambda_1 k^2+\frac{\lambda_2}{k^2}+\lambda_3.
\end{gather*}
By Proposition 2.1, eq. (\ref{CaseIb_kN}) admits the recursion operator
\begin{gather}
R[u]=D_x^2+2\left(\frac{k''}{k'}-\frac{k'}{k}\right)u_xD_x
+\left(\frac{k''}{k'}-\frac{k'}{k}\right)u_{xx}\notag\\[0.3cm]
\qquad
+\left(\frac{k'''}{k'}-3\frac{k''}{k}+\frac{(k')^2}{k^2}\right)u_{x}^2
+\frac{2}{3}\left(\lambda_1k^2+\frac{\lambda_2}{k^2}\right)\\[0.3cm]
\label{CaseIb_RkN}
\qquad
+u_xD_x^{-1}\left[\frac{(k')^2}{k^2}u_{xx}+\left(\frac{k'k''}{k^2}
-\frac{(k')^3}{k^3}\right)u_x^2
+\frac{2\lambda_1}{3}kk'-\frac{2\lambda_2}{3}\frac{k'}{k^3}\right].\notag
\end{gather}

\strut\hfill

\noindent
{\bf Example 2.4: i)} Let
\begin{gather*}
n(u)=0,\qquad m(u)=-\frac{\beta^2}{8}\\[0.3cm]
q(u)=\lambda_1e^{\beta u}+\lambda_2 e^{-\beta u}+\lambda_3,\qquad
\beta,\ \lambda_1,\ \lambda_2,\ \lambda_3\in\Re,
\end{gather*}
which satisfies system (\ref{gen_eq1_1}) -- (\ref{gen_eq1_2}). Then
(\ref{gen_eq1}) takes the form
\begin{gather}
\boxed{
\label{CaseI_e1}
u_t=u_{xxx}-\frac{\beta^2}{8}u_x^3+
\lambda_1e^{\beta u}u_x+\lambda_2 e^{-\beta u}u_x+\lambda_3u_x 
}
\end{gather}
and by Proposition 2.1, eq. (\ref{CaseI_e1}) admits the recursion operator
\begin{gather}
\label{CaseIe1_R}
R[u]=D_x^2-\frac{\beta^2}{4}u_x^2+
\frac{2\lambda_1}{3}e^{\beta u}+
\frac{2\lambda_2}{3}e^{-\beta u}\notag\\[0.3cm]
\qquad
+u_xD_x^{-1}\left[\frac{\beta^2}{4}u_{xx}
+\frac{\lambda_1\beta}{3}e^{\beta u}
-\frac{\lambda_2\beta}{3}e^{-\beta u}\right].
\end{gather}
This result is also given in \cite{fokas80}.
The next equation in the hierarchy is
\begin{gather}
u_t=u_{xxxxx}-\frac{5\beta^2}{8}u_x^2u_{xxx}-\frac{5\beta^2}{8}u_xu_{xx}^2
+\frac{3\beta^4}{128}u_x^5
+\frac{5\lambda_1^2}{6}e^{2\beta u}u_x
+\frac{4\lambda_1\lambda_2}{3}u_x
\notag\\[0.3cm]
\qquad
+\frac{5\lambda_1}{3}e^{\beta u}\left(
u_{xxx}+2\beta u_xu_{xx}+\frac{3\beta^2}{8}u_x^3\right)
+\frac{5\lambda_2}{3}e^{-\beta u}
\left(u_{xxx}-2\beta u_xu_{xx}
+\frac{3\beta^2}{8}u_x^3\right)\notag\\[0.3cm]
\qquad
+\frac{5\lambda_2^2}{6}e^{-2\beta u}u_x
+\lambda_3\left(
u_{xxx}-\frac{\beta^2}{8}u_x^3+\lambda_1e^{\beta u}u_x
+\lambda_2e^{-\beta u}u_x\right).
\label{ex1c_5i}
\end{gather}

\noindent
By the substitution
\begin{gather*}
u\to k(u),\qquad k\in {\cal C}^3,
\end{gather*}
eq. (\ref{CaseI_e1}) takes the form
\begin{gather}
\label{CaseIe1_k}
\boxed{
u_t=u_{xxx}
+3\frac{k''}{k'}u_xu_{xx}
+\left(\frac{k'''}{k'}-\frac{\beta^2}{8}(k')^2 \right)u_x^3
+\lambda_1e^{\beta k} u_x
+\lambda_2e^{-\beta k}u_x+\lambda_3u_x
}
\end{gather}
so that system (\ref{gen_eq1_1}) -- (\ref{gen_eq1_2}) admits the
solution
\begin{gather*}
n(u)=3\frac{k''}{k'},\qquad
m(u)=\frac{k'''}{k'}-\frac{\beta^2}{8}(k')^2\\[0.3cm]
q(u)=\lambda_1 e^{\beta k}+\lambda_2e^{-\beta k}+\lambda_3.
\end{gather*}
By Proposition 2.1, eq. (\ref{CaseIe1_k}) admits the recursion operator
\begin{gather}
\label{CaseIc_Rk1}
R[u]=D_x^2+2\frac{k''}{k'}u_xD_x+\frac{k''}{k'}u_{xx}
+\left(\frac{k'''}{k'}-\frac{\beta^2}{4}(k')^2\right)u_x^2
+\frac{2\lambda_1}{3}e^{\beta k}+\frac{2\lambda_2}{3}e^{-\beta k}
\notag\\[0.3cm]
\qquad
+u_xD_x^{-1}\left[
\frac{\beta^2}{4}\left((k')^2u_{xx}+k'k''u_x^2\right)+\frac{\beta\lambda_1}{3}
k'e^{\beta k}-\frac{\beta\lambda_2}{3}k'e^{-\beta k}\right].
\end{gather}

\strut\hfill

\noindent
{\bf Example 2.4: ii)}
Closely related to Example 2.4 i) is the solution
\begin{gather*}
n(u)=0,\qquad m(u)=\frac{\beta^2}{8}\\[0.3cm]
q(u)=\lambda_1\cos(\beta u)+\lambda_2 \sin(\beta u)+\lambda_3,\qquad
\beta,\ \lambda_1,\ \lambda_2,\ \lambda_3\in\Re,
\end{gather*}
which satisfies the system (\ref{gen_eq1_1}) -- (\ref{gen_eq1_2}). Then
(\ref{gen_eq1}) takes the form
\begin{gather}
\label{CaseI_e2}
\boxed{
u_t=u_{xxx}+\frac{\beta^2}{8}u_x^3+
\lambda_1\cos(\beta u)u_x+\lambda_2 \sin(\beta u)u_x+\lambda_3u_x 
}
\end{gather}
and by Proposition 2.1, eq. (\ref{CaseI_e2}) admits the recursion operator
\begin{gather}
\label{CaseIe2_R}
R[u]=D_x^2+\frac{\beta^2}{4}u_x^2+
\frac{2}{3}\lambda_1\cos(\beta u)+
\frac{2}{3}\lambda_2\sin(\beta u)\notag\\[0.3cm]
\qquad
-u_xD_x^{-1}\left[\frac{\beta^2}{4}u_{xx}
+\frac{\lambda_1\beta}{3}\sin(\beta u)
-\frac{\lambda_2\beta}{3}\cos(\beta u)\right].
\end{gather}
The next equation in this hierarchy is
\begin{gather}
u_t=u_{xxxxx}+\frac{5\beta^2}{8}u_x^2u_{xxx}
+\frac{5\beta^2}{8}u_xu_{xx}^2
+\frac{3\beta^4}{128}u_x^5
+\frac{5\lambda_2^2}{6}\sin^2(\beta u)u_x\notag\\[0.3cm]
\qquad
+\frac{5\lambda_1\lambda_2}{3}\cos(\beta u)\sin(\beta u)u_x
+\frac{5\lambda_1^2}{6}\cos^2(\beta u)u_x
-\frac{\lambda_1^2}{6}u_x-\frac{\lambda_2^2}{6}u_x
\notag\\[0.3cm]
\qquad
+\frac{5\lambda_1}{3}\left(
\cos (\beta u)u_{xxx}-2\beta\sin(\beta u) u_xu_{xx}
-\frac{3\beta^2}{8}\cos(\beta u)u_x^3\right)\notag\\[0.3cm]
\qquad
+\frac{5\lambda_2}{3}
\left(\sin(\beta u)u_{xxx}+2\beta \cos(\beta u)u_xu_{xx}
-\frac{3\beta^2}{8}\sin(\beta u)u_x^3\right)\notag\\[0.3cm]
\qquad
+\lambda_3\left(
u_{xxx}+\frac{\beta^2}{8}u_x^3+\lambda_1\cos(\beta u)u_x
+\lambda_2\sin(\beta u)u_x\right).
\label{ex1c_5ii}
\end{gather}
By the substitution
\begin{gather*}
u\to k(u),\qquad k\in {\cal C}^3,
\end{gather*}
eq. (\ref{CaseI_e2}) takes the form
\beg
\boxed{
\label{CaseIe2_k}
\ba{l}
\displaystyle{
u_t=u_{xxx}
+3\frac{k''}{k'}u_xu_{xx}
+\left(\frac{k'''}{k'}+\frac{\beta^2}{8}(k')^2 \right)u_x^3}\\
[0.5cm]
\displaystyle{
\qquad+\lambda_1 \cos(\beta k) u_x
+\lambda_2\sin(\beta k)u_x+\lambda_3u_x}
\ea
}
\eeq
so that system (\ref{gen_eq1_1}) -- (\ref{gen_eq1_2}) admits the
solution
\begin{gather*}
n(u)=3\frac{k''}{k'},\qquad
m(u)=\frac{k'''}{k'}+\frac{\beta^2}{8}(k')^2\\[0.3cm]
q(u)=\lambda_1 \cos(\beta k)+\lambda_2\sin(\beta k)+\lambda_3.
\end{gather*}
By Proposition 2.1, eq. (\ref{CaseIe2_k}) admits the recursion operator
\begin{gather}
\label{CaseIc_Rk2}
R[u]=D_x^2+2\frac{k''}{k'}u_xD_x+\frac{k''}{k'}u_{xx}
+\left(\frac{k'''}{k'}+\frac{\beta^2}{4}(k')^2\right)u_x^2\notag\\[0.3cm]
\qquad 
+\frac{2\lambda_1}{3}\cos(\beta k)+\frac{2\lambda_2}{3}\sin(\beta k)
\notag\\[0.3cm]
\qquad
-u_xD_x^{-1}\left[
\frac{\beta^2}{4}\left((k')^2u_{xx}+k'k''u_x^2\right)+\frac{\beta\lambda_1}{3}
k'\sin(\beta k)-\frac{\beta\lambda_2}{3}k'\cos(\beta k)\right].
\end{gather}

\strut\hfill

Our last example is on the well known KdV and modified KdV
equation, which should obviously be contained in Proposition 2.1.
The recursion operators for these equations are well known (see for
example \cite{fokas80})

\strut\hfill

\noindent
{\bf Example 2.5:} Let
\begin{gather*}
m(u)=0,\qquad n(u)=0\\[0.3cm]
q(u)=\lambda_1u^2+\lambda_2u+\lambda_3,\qquad
\lambda_1,\ \lambda_2,\ \lambda_3\in\Re,
\end{gather*}
which satisfies system (\ref{gen_eq1_1}) -- (\ref{gen_eq1_2}). Then
(\ref{gen_eq1}) takes the form
\begin{gather}
\label{CaseI_f}
\boxed{
u_t=u_{xxx}+\lambda_1 u^2u_x+\lambda_2uu_x+\lambda_3u_x
}
\end{gather}
and admits the recursion operator
\begin{gather}
\label{CaseIf_R}
R[u]=D_x^2+
\frac{2\lambda_1}{3}u^2+
\frac{2\lambda_2}{3}u
+\frac{1}{3}u_xD_x^{-1}
\left[2\lambda_1u+\lambda_2\right].
\end{gather}
With $\lambda_1=0$, eq. (\ref{CaseI_f}) is the
KdV equation and for $\lambda_2=0$
eq. (\ref{CaseI_f}) is known as the
modified KdV equation. This result
is also given in \cite{fokas80}.
The next equation in this hierarchy is
\begin{gather}
u_t=u_{xxxxx}+\frac{5\lambda_1}{3}\left(u^2u_{xxx}+4uu_xu_{xx}+u_x^3\right)
+\frac{5\lambda_2}{3}\left(uu_{xxx}+2u_xu_{xx}\right)\notag\\[0.3cm]
\qquad
+\frac{5\lambda_1^2}{6}u^4u_x
+\frac{5\lambda_1\lambda_2}{3}u^3u_x
+\frac{5\lambda_2^2}{6}u^2u_x
+\lambda_3\left(u_{xxx}+\lambda_1u^2u_x+\lambda_2uu_x\right).
\label{ex1d_5}
\end{gather}
By the substitution
\begin{gather*}
u\rightarrow k(u),\qquad k\in {\cal C}^3
\end{gather*}
eq. (\ref{CaseI_f}) takes the form
\begin{gather}
\label{CaseI_fk}
\boxed{
u_t=u_{xxx}+3\frac{k''}{k'}u_xu_{xx}+\frac{k'''}{k'}u_x^3
+\lambda_1 k^2u_x+\lambda_2ku_x+\lambda_3u_x
}
\end{gather}
so that system (\ref{gen_eq1_1}) -- (\ref{gen_eq1_2}) admits the solution
\begin{gather*}
m(u)=\frac{k'''}{k'},\qquad n(u)=3\frac{k''}{k'}\\[0.3cm]
q(u)=\lambda_1k^2+\lambda_2k+\lambda_3,\qquad
\lambda_1,\ \lambda_2,\ \lambda_3\in\Re.
\end{gather*}
By Proposition 2.1, eq. (\ref{CaseI_fk}) admits the recursion operator
\begin{gather}
\label{CaseIfk_R}
R[u]=D_x^2+2\frac{k''}{k'}u_xD_x
+\frac{k''}{k'}u_{xx}+\frac{k'''}{k'}u_x^2
+\frac{2}{3}\left(\lambda_1 k^2+\lambda_2 k\right)
\notag\\[0.3cm]
\qquad
+\frac{1}{3}u_xD^{-1}_x\left[2\lambda_1kk'+\lambda_2 k'\right].
\end{gather}

\section{Recursion operators for the quasilinear equation (\ref{gen_quasi})}
The recursion operator Ansatz (\ref{gen_R}) for (\ref{gen_quasi})
leads to 

\strut\hfill

\noindent
{\bf Proposition 3.1} {\it
Evolution equations of the form
\begin{gather}
u_t=u^3u_{xxx}+n(u)u_xu_{xx}+m(u)u_x^3+ r(u)u_{xx} +p(u)u_x^2+
q(u)u_x+s(u),
\notag
\end{gather}
which admit recursion operators of the form (\ref{gen_R}), viz.
\begin{gather*}
R[u]=GD_x^2+QD_x+H+\sum_jI_jD_x^{-1}J_j
\end{gather*}
and which are not linearisable by a $x$-generalised hodograph transformation, 
are exhausted by the following two cases:\n
\begin{enumerate} 
\item[{\bf I)}] The equation
\begin{gather}
\label{gen_quasi1}
\boxed{
u_t=u^3u_{xxx}+3u^2u_xu_{xx}+\lambda_1u^3u_x+\lambda_2u^2u_x+\lambda_3u_x}
\quad
\lambda_j\in \Re
\end{gather}
admits the recursion operator
\begin{gather}
\label{quasi_recI}
R[u]=u^2D_x^2+uu_xD_x+2uu_{xx}+\lambda_1u^2
+\frac{4\lambda_2}{3}u\notag\\[0.3cm]
\qquad+\left(
u^3u_{xxx}+3u^2u_xu_{xx}+\lambda_1u^3u_x+\lambda_2u^2u_x
\right)D_x^{-1}u^{-2}
-\frac{\lambda_2}{3}u_xD_x^{-1}.
\end{gather}
\item[{\bf II)}] The equation
\begin{gather}
\label{gen_quasi2}
\boxed{
u_t=u^3u_{xxx}+\lambda_1u^3u_x+\lambda_2u^{-1}u_x+\lambda_3u_x}\quad
\lambda_j\in \Re
\end{gather}
admits the recursion operator
\begin{gather}
\label{quasi_recII}
R[u]=u^2D_x^2-uu_xD_x+uu_{xx}+\lambda_1u^2
+\frac{\lambda_2}{3}u^{-2}\notag\\[0.3cm]
\qquad +\left(
u^3u_{xxx}+\lambda_1u^3u_x+\lambda_2u^{-1}u_x
\right)D_x^{-1}u^{-2}-\frac{4\lambda_2}{3}u_xD_x^{-1}u^{-3}.
\end{gather}
\end{enumerate}
}

\strut\hfill

\noindent
To prove Proposition 3.1 one needs to
verify commutator realtion
(\ref{commutator}) for eq. (\ref{gen_quasi}) and the recursion operator
Ansatz (\ref{gen_R}). This is done by direct calculations, so we do
not show the details here.

\strut\hfill

We apply the recursion operators of Proposition 3.1 on the
$t$-translation symmetry, i.e. $R[u]u_t$,
to obtain the next equation in the hierarchies.
For eq. (\ref{gen_quasi1}) the next equation in the
hierarchy is
\begin{gather}
\label{gen5_quasi1}
u_t=u^5u_{xxxxx}+10u^4u_xu_{xxxx}
+15u^4u_{xx}u_{xxx}+25u^3u_x^2u_{xxx}+30u^3u_xu_{xx}^2\notag\\[0.3cm]
\qquad +15u^2u_x^3u_{xx}
+\frac{5\lambda_1}{2}\left(u^5u_{xxx}+7u^4u_xu_{xx}
+4u^3u_x^3\right)\notag\\[0.3cm]
\qquad+\frac{10\lambda_2}{3}\left(u^4u_{xxx}+5u^3u_xu_{xx}
+\frac{3}{2}u^2u_x^3\right)
\notag\\[0.3cm]
\qquad
+\frac{3\lambda_1^2}{2}u^5u_x+\frac{15\lambda_1\lambda_2}{4}u^4u_x
+\frac{20\lambda_2^2}{9}u^3u_x
\end{gather}
and for eq. (\ref{gen_quasi2}) the next equation in the
hierarchy is
\begin{gather}
\label{gen5_quasi2}
u_t=u^5u_{xxxxx}+5u^4u_xu_{xxxx}
+5u^4u_{xx}u_{xxx}+\frac{5}{2}u^3u_x^2u_{xxx}\notag\\
\qquad
+\frac{5\lambda_1}{2}\left(u^5u_{xxx}+4u^4u_xu_{xx}+u^3u_x^3\right)
+\frac{5\lambda_2}{6}\left(uu_{xxx}-4u_xu_{xx}
+3u^{-1}u_x^3\right)\notag\\[0.3cm]
\qquad
+\frac{3\lambda_1^2}{2}u^5u_x+\frac{5\lambda_2^2}{18}u^{-3}u_x.
\end{gather}

\strut\hfill

We now use the substitution
\bg
u\to k(u)
\ed
for the two equations of Proposition 3.1.
For eq. (\ref{gen_quasi1}) we obtain
\beg
\boxed{
\label{gen_quasi1k}
\ba{l}
\displaystyle{
u_t=k^3u_{xxx}+3\left(k^3\frac{k''}{k'}+k^2k'\right)u_xu_{xx}+
\left(k^3\frac{k'''}{k'}+3k^2k''\right)u_x^3}\\
[0.5cm]
\displaystyle{
\qquad +\lambda_1k^3u_x+\lambda_2 k^2u_x+\lambda_3u_x}
\ea
}
\eeq
with the recursion operator
\begin{gather}
R[u]=k^2D_x^2+\left(2k^2\frac{k''}{k'}+kk'\right)u_xD_x
+\left(k^2\frac{k''}{k'}+2kk'\right)u_{xx}
+\lambda_1k^2
+\frac{4\lambda_2}{3}k\notag\\[0.3cm]
\qquad +\left(k^2\frac{k'''}{k'}+3kk''\right)u_x^2
+\left\{
k^3u_{xxx}+3\left(k^3\frac{k''}{k'}+k^2k'\right)u_xu_{xx}\right.\notag
\\[0.3cm]
\qquad \left.+
\left(k^3\frac{k'''}{k'}+3k^2k''\right)u_x^3
+\lambda_1k^3u_x+\lambda_2 k^2u_x
\right\}D_x^{-1}\frac{k'}{k^2}
-\frac{\lambda_2}{3}u_xD_x^{-1}k'.
\label{quasi_recIk}
\end{gather}
For eq. (\ref{gen_quasi2}) 
we obtain
\begin{gather}
\label{gen_quasi2k}
\boxed{
u_t=k^3u_{xxx}+3k^3\frac{k''}{k'}u_xu_{xx}+
k^3\frac{k'''}{k'}u_x^3
+\lambda_1k^3u_x+\frac{\lambda_2}{k}u_x+\lambda_3u_x
}
\end{gather}
and its recursion operator is
\begin{gather}
\label{quasi_recIIk}
R[u]=k^2D_x^2+\left(2k^2\frac{k''}{k'}-kk'\right)u_xD_x
+\left(k^2\frac{k''}{k'}+kk'\right)u_{xx}
+\lambda_1k^2\notag\\[0.3cm]
\qquad +k^2\frac{k'''}{k'}u_x^2
+\frac{\lambda_2}{3}\frac{1}{k^2}
-\frac{4\lambda_2}{3}u_xD_x^{-1}\frac{k'}{k^3}\notag\\[0.3cm]
\qquad+\left(
k^3u_{xxx}+3k^3\frac{k''}{k'}u_xu_{xx}
+k^3\frac{k'''}{k'}u_x^3
+\lambda_1 k^3u_x+\frac{\lambda_2}{k}u_x\right)D_x^{-1}\frac{k'}{k^2}.
\end{gather}

\section{Transformations between equations}

\subsection{From semilinear to quasilinear equations}

We make use of the $x$-generalised hodograph transformation
(\ref{n_hodo}) to transform those semilinear integrable equations,
given by Proposition 2.1, to quasilinear equations
within the class of autonomous evolution equations
\begin{gather*}
u_t=F(u,u_x,u_{xx},u_{xxx}).
\end{gather*}

Equation (\ref{gen_eq3k}) given by Proposition 2.1 Case I, namely
\begin{gather*}
u_t=u_{xxx}+3\frac{k''}{k'}u_xu_{xx}+\left(\frac{k'''}{k'}+\lambda_1
(k')^2\right)u_x^3+\lambda_2 k' u_x^2+ \lambda_3 u_x+\frac{\lambda_4}{k'},
\end{gather*}
is transformed by the $x$-generalised hodograph transformation
\begin{gather}
\label{3_hodo2}
_3\mbox{\bf H}:\left\{\ba{l}
\displaystyle{dx(\tilde x,\tilde t)
=
\tilde k^{-1}d\tilde x}\\[3mm]
\hphantom{dx(\tilde x,\tilde t)}
\displaystyle{
+\left\{-\tilde k\tilde k' \tilde u_{\tilde x\tilde x}
-[(\tilde k')^2+\tilde k\tilde k'']\tilde u_{\tilde x}^2
-\lambda_1 \tilde k^2-\lambda_2\tilde k-
\lambda_4\tilde k^{-1}-\lambda_3\right\}d\tilde t}
\\[3mm]
dt(\tilde x,\tilde t)=d\tilde t \\[3mm]
k(u)=\tilde x,
\ea\right.
\end{gather}
to the quasilinear equation (\ref{gen_quasi1k}), namely
\begin{gather*}
\tilde u_{\tilde t}
=\tilde k^3\tilde u_{\tilde x\tilde x\tilde x}
+3\left(\tilde k^3\frac{\tilde k''}{\tilde k'}+\tilde k^2\tilde
k'\right)
\tilde u_{\tilde x}\tilde u_{\tilde x\tilde x}+
\left(\tilde k^3\frac{\tilde k'''}{\tilde k'}+3\tilde k^2\tilde
k''\right)
\tilde u_{\tilde x}^3\notag\\[0.3cm]
\label{tilde_2}
\qquad
+2\lambda_1\tilde k^3\tilde u_{\tilde x}+\lambda_2 \tilde k^2\tilde
u_{\tilde x}-\lambda_4 \tilde u_{\tilde x}.
\end{gather*}
We recall that eq. (\ref{gen_quasi1k}) admits the
recursion operator (\ref{quasi_recIk}).

\strut\hfill

The only integrable equation of Proposition 2.1, Case II,
that can be transformed by a $x$-generalised hodograph
transformation within the class of autonomous equations,
is the KdV equation (\ref{CaseI_fk})
with $\lambda_1=0$, i.e.
\begin{gather}
\label{CaseIlam2_0}
u_t=u_{xxx}+3\frac{k''}{k'}u_xu_{xx}+\frac{k'''}{k'}u_x^3
+\lambda_2ku_x+\lambda_3u_x.
\end{gather}
By the the $x$-generalised hodograph transformation
\begin{gather}
\label{3_hodo1}
_3\mbox{\bf H}:\left\{\ba{l}
\displaystyle{dx(\tilde x,\tilde t)
=
\tilde k^{-1}d\tilde x}\\[3mm]
\hphantom{dx(\tilde x,\tilde t)}
\displaystyle{
+\left\{-\tilde k\tilde k' \tilde u_{\tilde x\tilde x}
-\left((\tilde k')^2+\tilde k\tilde k''\right)\tilde u_{\tilde x}^2
-\lambda_2 \tilde x-\lambda_3
\right\}d\tilde t}
\\[3mm]
dt(\tilde x,\tilde t)=d\tilde t \\[3mm]
k(u)=\tilde x,
\ea\right.
\end{gather}
we obtain for (\ref{CaseIlam2_0})
the following quasilinear equation
\begin{gather}
\label{quasi1_tilde}
\boxed{
\tilde u_{\tilde t}=
\tilde k^3\tilde u_{\tilde x\tilde x\tilde x}
+3\left(\tilde k^3\frac{\tilde k''}{\tilde k'}
+\tilde k^2\tilde k'\right)\tilde u_{\tilde x}
\tilde u_{\tilde x\tilde x}+
\left(\tilde k^3\frac{\tilde k'''}{\tilde k'}+3\tilde k^2\tilde
k''\right)u_x^3
+\lambda_2 \frac{\tilde k^2}{\tilde k'}
}
\end{gather}
Equation (\ref{quasi1_tilde}) does not admit a recursion operator
of the form (\ref{gen_R}).

It is interesting to note that both (3.3) with
$\lambda_1=\lambda_2=\lambda_3=0$, i.e. the well known
{\it Harry-Dym equation}
\begin{gather}
\label{H-D}
u_t=u^3u_{xxx},
\end{gather}
and (2.1) with $\lambda_2=\lambda_4=0$, i.e the potential modified KdV
equation
\begin{gather}
\label{PmKdV}
\tilde u_{\tilde t}=\tilde u_{\tilde x\tilde x\tilde x}+\lambda_1
\tilde u_{\tilde x}^3+\lambda_3\tilde u_{\tilde x},
\end{gather}
can be transformed to eq. (\ref{quasi1_tilde}) by a $x$-generalised
hodograph transformation. The composition of those two $x$-generalised
hodograph transformations results in the following transformation
between eq. (\ref{H-D}) and eq. (\ref{PmKdV}):
\begin{gather}
_3\mbox{\bf H}:\left\{\ba{l}
\displaystyle{dx(\tilde x,\tilde t)
=
\frac{1}{\alpha}e^{\alpha \tilde u}d\tilde x}
\displaystyle{
+\frac{1}{\alpha}e^{\alpha \tilde u}
\left\{\alpha \tilde u_{\tilde x\tilde x}
+\lambda_1\tilde u^2_{\tilde x}+\lambda_3
\right\}d\tilde t}
\\[3mm]
dt(\tilde x,\tilde t)=d\tilde t \\[3mm]
\displaystyle{u(x,t)=\frac{1}{\alpha}e^{\alpha \tilde u},\qquad
\alpha^2=-2\lambda_1,\quad \alpha=\mbox{nonzero constant.}  } 
\ea\right.
\end{gather}
We remark that this relation between the Harry-Dym and the modified
KdV was also given in \cite{kawamoto}.

\subsection{Potential forms for semilinear equations of Proposition 2.1}

We now consider eq.(\ref{CaseI_b}) of Example 2.2 and
eq. (\ref{CaseI_e1}) of Example 2.4 i).
Our aim is to
perform the following change of coordinates successively:
\begin{gather}
\label{pot}
(x,t,u(x,t)) \mapsto (x,t, v(x,t))\quad{
\mbox{Potential Form (PF) with }}\ \ v_x=h(u)\\[0.3cm]
\label{HT}
(x,t,v(x,t))\mapsto (X,t,V(X,t))
\quad{\mbox{Hodograph Transformation (HT)}}\\[0.3cm]
\label{npot}
(X,t,V(X,t))\mapsto (X,t,W(X,t))\quad {\mbox{with}}\ \ W=V_X \\[0.3cm]
\label{xGHT}
(X,t,W(X,t))\mapsto (\tilde x,\tilde t,\tilde u(\tilde x,\tilde t))
\quad{\mbox{$x$-Generalised Hodograph Transformation}}
\end{gather}
The hodograph transformation (\ref{HT})
has the following explicit form:
\begin{gather}
\label{TheHT}
X=v(x,t),\qquad T=t,\qquad  V(X,t)=x.
\end{gather}

\strut\hfill

\noindent
{\bf 4.2.1}\ \  We consider eq. (\ref{CaseI_b}), i.e.
\begin{gather*}
u_t=u_{xxx}+3u^{-1}u_xu_{xx}+ \lambda_1u^4u_x+\lambda_2 u^2u_x+\lambda_3u_x, 
\end{gather*}
which admits two potential forms. Those are discussed separately under
4.2.1 i) and 4.2.1 ii) below:

\strut\hfill

\noindent
{\bf 4.2.1 i)} For eq. (\ref{CaseI_b}) we consider the change of
coordinates (\ref{pot}) and set
\begin{gather*}
v_x=u^4
\end{gather*}
The potential equation is
\begin{gather}
\boxed{
\label{CaseI_bpot}
v_t=v_{xxx}-\frac{3}{4}v_x^{-1}v_{xx}^2+\frac{1}{2}\lambda_1 v_x^2
+\frac{2}{3}\lambda_2v_x^{3/2}+\lambda_3 v_x+C
}\quad C\in \Re.
\end{gather}
Under the hodograph transformation (\ref{TheHT}), eq. (\ref{CaseI_bpot}) takes the form
\begin{gather}
\label{CaseI_HT}
V_t=V_X^{-3}V_{XXX}-\frac{9}{4}V_X^{-4}V_{XX}^2-\frac{1}{2}\lambda_1V_X^{-1}
-\frac{2}{3}\lambda_2V_X^{-1/2}-CV_X-\lambda_3.
\end{gather}
We remark that if we now apply the $x$-generalised hodograph
transformation (\ref{n_hodo}) on (\ref{CaseI_HT}) then
we obtain (\ref{CaseI_b}), since
\begin{gather}
(\mbox{$x$-Generalised Hodograph Transformation})^{-1}\ =\
\mbox{HT\ $\circ$\ PF}.
\end{gather}
We now apply (\ref{npot}), which transforms eq. (\ref{CaseI_HT}) to
\beg
\label{Eq_W1}
\ba{l}
\displaystyle{
W_t=W^{-3}W_{XXX}-\frac{15}{2}W^{-4}W_XW_{XX}+9W^{-5}W_X^3
+\frac{1}{2}\lambda_1 W^{-2}W_X}\\
\displaystyle{
\qquad
+\frac{1}{3}\lambda_2W^{-3/2}W_X-CW_X}.
\ea
\eeq
It is worth to note that, with $\lambda_1=0$, eq. (\ref{Eq_W1})
transforms to (\ref{quasi1_tilde}) by the $x$-generalised hodograph
transformation
\begin{gather}
\label{3_hodoW}
_3\mbox{\bf H}:\left\{\ba{l}
\displaystyle{dX(\tilde x,\tilde t)
=
\tilde x^2\tilde k^{-1}d\tilde x}\\[3mm]
\hphantom{dx(\tilde x,\tilde t)}
\displaystyle{
+\left\{-\tilde k^2 +2\tilde x\tilde k\tilde k' \tilde u_{\tilde x}
-\tilde x^2[(\tilde k')^2+\tilde k\tilde k'']\tilde u_{\tilde x}^2
-\tilde x^2\tilde k\tilde k'\tilde u_{\tilde x\tilde x}
+\frac{\lambda_2}{3}\tilde x^3+C\right\}d\tilde t}
\\[3mm]
dt(\tilde x,\tilde t)=d\tilde t \\[3mm]
W(X,t)=\tilde x^{-2}.
\ea\right.
\end{gather}
We recall that (\ref{quasi1_tilde})
can be transformed to
the KdV equation (see (\ref{CaseIlam2_0}) and (\ref{quasi1_tilde})).

\strut\hfill

\noindent
{\bf 4.2.1 ii)} For the second potential form of (\ref{CaseI_b})
we set
\begin{gather*}
v_x=u^2,
\end{gather*}
which leads to
\begin{gather}
\label{CaseI_bpot2}
v_t=v_{xxx}+\frac{1}{3}\lambda_1v_x^3+\frac{1}{2}\lambda_2v_x^2
+\lambda_3v_x+C.
\end{gather}
Note that by a simple change of constants
eq.(\ref{CaseI_bpot2}) is equivalent to
eq. (\ref{gen_eq3}) of Proposition 2.1.
Under the hodograph transformation (\ref{TheHT}), eq. (\ref{CaseI_bpot2}) takes the form
\begin{gather}
\label{CaseI_HT2}
V_t=V_X^{-3}V_{XXX}-3V_X^{-4}V_{XX}^2-\frac{1}{3}\lambda_1V_X^{-2}
-\frac{1}{2}\lambda_2V_X^{-1}-CV_X-\lambda_3.
\end{gather}
We now apply (\ref{npot}), so eq. (\ref{CaseI_HT2}) becomes
\beg
\label{Eq_W2}
\ba{l}
\displaystyle{
W_t=W^{-3}W_{XXX}-9W^{-4}W_XW_{XX}+12W^{-5}W_X^3
+\frac{2}{3}\lambda_1 W^{-3}W_X}\\[0.3cm]
\displaystyle{
\qquad
+\frac{1}{2}\lambda_2W^{-2}W_X-CW_X}
\ea
\eeq
Equation (\ref{Eq_W2}) is equivalent to
(\ref{gen_quasi1}) with $C=\lambda_3$ and
\begin{gather*}
u(x,t)=W^{-1}(X,t),\qquad X=x.
\end{gather*}

\strut\hfill

\noindent
{\bf 4.2.2}\ \ We now consider eq. (\ref{CaseI_e1}) of
Example {\bf 2.4 i}, i.e.
\begin{gather*}
u_t=u_{xxx}-\frac{\beta^2}{8}u_x^3+
\lambda_1e^{\beta u}u_x+\lambda_2 e^{-\beta u}u_x+\lambda_3u_x. 
\end{gather*}
For the change of coordinates (\ref{pot}) we set
\begin{gather*}
v_x=\exp\left\{\frac{\beta u}{2}\right\}
\end{gather*}
and obtain the potential equation
\begin{gather}
\boxed{
\label{CaseI_epot}
v_t=v_{xxx}-\frac{3}{2}v_x^{-1}v_{xx}^2+\frac{1}{3}\lambda_1 v_x^3
-\lambda_2v_x^{-1}+\lambda_3 v_x+C
}\quad C\in \Re.
\end{gather}
We remark that, with
\begin{gather*}
\lambda_1=\lambda_2=\lambda_3=C=0,
\end{gather*}
eq. (\ref{CaseI_epot}) is known as the {\it Kirchever-Novikov equation}
\cite{Dorfman}
and admits the recursion operator \cite{Wang}
\begin{gather}
R[v]=D_x^2-2v_x^{-1}v_{xx}D_x
+\left(v_x^{-1}v_{xxx}-v_x^{-2}v_{xx}^{2}\right)
\notag\\[0.3cm]
\label{KN}
\qquad
-v_xD_x^{-1}\left[3v_x^{-4}v_{xx}^3-4v_x^{-3}v_{xx}v_{xxx}+v_x^{-2}v_{xxxx}
\right].
\end{gather}
Under the hodograph transformation (\ref{TheHT}), eq. (\ref{CaseI_epot}) takes the form
\begin{gather}
\label{CaseIe_HT}
V_t=V_X^{-3}V_{XXX}-\frac{3}{2}V_X^{-4}V_{XX}^2-\frac{1}{3}\lambda_1V_X^{-2}
+\lambda_2V_X^{2}-CV_X-\lambda_3.
\end{gather}
We now apply (\ref{npot}), which transforms eq. (\ref{CaseIe_HT})
to
\begin{gather}
W_t=W^{-3}W_{XXX}-6W^{-4}W_XW_{XX}+6W^{-5}W_X^3
+\frac{2}{3}\lambda_1 W^{-3}W_X\notag\\[0.3cm]
\label{Eq_We}
\qquad
+\lambda_2WW_X-CW_X.
\end{gather}
Equation (\ref{Eq_We}) is equivalent to
(\ref{gen_quasi2}) with $C=\lambda_3$ and
\begin{gather*}
u(x,t)=W^{-1}(X,t),\qquad X=x.
\end{gather*}

\subsection{Potential forms for quasilinear equations of Proposition 3.1}

We now consider the quasilinear equations of Proposition 2, i.e.
equations (\ref{gen_quasi1}) and (\ref{gen_quasi2}). As in the
previous subsection we apply the sequence of coordinate transformations
(\ref{pot})--(\ref{xGHT}).

\strut\hfill

\noindent
{\bf 4.3.1}\ \ Consider eq. (\ref{gen_quasi1}), i.e.
\begin{gather*}
u_t=u^3u_{xxx}+3u^2u_xu_{xx}+\lambda_1u^3u_x+\lambda_2u^2u_x+\lambda_3u_x
\quad
\lambda_j\in \Re
\end{gather*}
which admits two potential forms. Those are discussed separately under
4.3.1 i) and 4.3.1 ii) below:

\strut\hfill

\noindent
{\bf 4.3.1 i)} For eq. (\ref{gen_quasi1}) we consider the change of
coordinates (\ref{pot}) and set
\begin{gather*}
v_x=u^{-1}
\end{gather*}
The potential equation is
\begin{gather}
\boxed{
\label{quasi1_bpot}
v_t=v_x^{-3}v_{xxx}-3v_x^{-4}v_{xx}^2-\frac{\lambda_1}{2} v_x^{-2}
-\lambda_2v_x^{-1}+\lambda_3 v_x+C
}\quad C\in \Re.
\end{gather}
Under the hodograph transformation (\ref{TheHT}), eq.
(\ref{quasi1_bpot}) takes the form
\begin{gather}
\label{quasi1_HT}
V_t=V_{XXX}+\frac{\lambda_1}{2}V_X^{3}+\lambda_2V_X^{2}-CV_X-\lambda_3
\end{gather}
which is just eq. (\ref{gen_eq3}).

\strut\hfill

\noindent
{\bf 4.3.1 ii)} To obatin the second potential form for (\ref{gen_quasi1})
we set
\begin{gather*}
v_x=u.
\end{gather*}
The potential equation is
\begin{gather}
\boxed{
\label{quasi1_bpot2}
v_t=v_x^{3}v_{xxx}+\frac{\lambda_1}{4}v_x^{4}+\frac{\lambda_2}{3} v_x^{3}
+\lambda_3 v_x+C
}\quad C\in \Re.
\end{gather}
Under the hodograph transformation (\ref{TheHT}), eq.
(\ref{quasi1_bpot2}) takes the form
\begin{gather}
\label{quasi1_HT2}
V_t=V_{X}^{-6}V_{XXX}-3V_X^{-7}V_{XX}^2
-\frac{\lambda_1}{4}V_X^{-3}
-\frac{\lambda_2}{3}V_X^{-2}-CV_X-\lambda_3.
\end{gather}
We now apply (\ref{npot}), which transforms eq. (\ref{quasi1_HT2})
to
\beg
\label{quasi1_W2}
\ba{l}
\displaystyle{
W_t=W^{-6}W_{XXX}-12W^{-7}W_XW_{XX}+21W^{-8}W_X^3}\\
[0.3cm]
\qquad
\displaystyle{
+\frac{3\lambda_1}{4}W^{-4}W_X
+\frac{2\lambda_2}{3}W^{-3}W_X-CW_X}.
\ea
\eeq
Note that, with $\lambda_1=0$, eq. (\ref{quasi1_W2})
can be transformed to (\ref{quasi1_tilde}) by the
$x$-generalised hodograph transformation
\begin{gather*}
\label{3_hodoWquasi}
_3\mbox{\bf H}:\left\{\ba{l}
\displaystyle{dX(\tilde x,\tilde t)
=
\frac{1}{4}\tilde x^2\tilde k^{-1}d\tilde x}\\[3mm]
\hphantom{dx(\tilde x}
\displaystyle{
+\left\{-\frac{1}{4}\tilde k^2 +\frac{1}{2}\tilde x\tilde k
\tilde k' \tilde u_{\tilde x}
-\frac{\tilde x^2}{4}[(\tilde k')^2+\tilde k\tilde k'']\tilde u_{\tilde x}^2
-\frac{\tilde x^2}{4}\tilde k\tilde k'\tilde u_{\tilde x\tilde x}
-\frac{\lambda_2}{12}\tilde x^3+C\right\}d\tilde t}
\\[3mm]
dt(\tilde x,\tilde t)=d\tilde t \\[3mm]
W(X,t)=2\tilde x^{-1}.
\ea\right.
\end{gather*}
We recall that (\ref{quasi1_tilde}) can be transformed to the KdV
equation (\ref{CaseIlam2_0}).

\strut\hfill

\noindent
{\bf 4.3.2}\ \  We now consider (\ref{gen_quasi2}), i.e.
\begin{gather*}
u_t=u^3u_{xxx}+\lambda_1u^3u_x+\lambda_2u^{-1}u_x+\lambda_3u_x,\quad
\lambda_j\in \Re
\end{gather*}
which admits two potential forms. Those are discussed separately under
4.3.2 i) and 4.3.2 ii) below:

\strut\hfill

\noindent
{\bf 4.3.2 i)} For eq. (\ref{gen_quasi2}) we consider the change of
coordinates (\ref{pot}) and set
\begin{gather*}
v_x=u^{-1}.
\end{gather*}
The potential equation is
\begin{gather}
\boxed{
\label{quasi2_bpot}
v_t=v_x^{-3}v_{xxx}-\frac{3}{2}v_x^{-4}v_{xx}^2-\frac{\lambda_1}{2} v_x^{-2}
+\frac{\lambda_2}{2}v_x^{2}+\lambda_3 v_x+C
}\quad C\in \Re.
\end{gather}
Under the hodograph transformation (\ref{TheHT}), eq.
(\ref{quasi2_bpot}) takes the form
\begin{gather}
\label{quasi2_HT}
V_t=V_{XXX}-\frac{3}{2}V_X^{-1}V_{XX}^2+\frac{\lambda_1}{2}V_{X}^3
-\frac{\lambda_2}{2}V_X^{-1}-CV_X-\lambda_3.
\end{gather}
We now apply (\ref{npot}), by which eq. (\ref{quasi2_HT}) becomes
\begin{gather}
W_t=W_{XXX}-3W^{-1}W_XW_{XX}
+\frac{3}{2}W^{-2}W_X^3
+\frac{3}{2}\lambda_1 W^2W_X\notag\\
\label{quasi2_W1}
\qquad+\frac{1}{2}\lambda_2W^{-2}W_X-CW_X.
\end{gather}
Equation (\ref{quasi2_W1}) is just eq.(\ref{CaseI_bN}) of Example 2.3.

\strut\hfill

\noindent
{\bf 4.3.2 ii)} For the second potential form of
eq. (\ref{gen_quasi2}) we set
\begin{gather*}
v_x=u^{-2}.
\end{gather*}
The potential equation is
\begin{gather}
\boxed{
\label{quasi2_bpot2}
v_t=v_x^{-3/2}v_{xxx}-\frac{3}{2}v_x^{-5/2}v_{xx}^2
-2\lambda_1 v_x^{-1/2}
+\frac{2\lambda_2}{3}v_x^{3/2}
+\lambda_3 v_x+C
}\quad C\in \Re.
\end{gather}
Equation (\ref{quasi2_bpot2}) is a slight generalisation of the
{\it Cavalcante-Tenenblat equation} \cite{Cav-Ten} and
admits the recursion operator
\begin{gather}
\label{CT_R}
R[v]=v_x^{-1}D_x^2-\frac{3}{2}v_x^{-2}v_{xx}D_x
-\frac{1}{2}v_x^{-2}v_{xxx}+\frac{3}{4}v_x^{-3}v_{xx}^2
+\lambda_1v_x^{-1}+\frac{\lambda_2}{3}v_x\notag\\[0.3cm]
\qquad
-\frac{1}{4}\left(v_x^{-3/2}v_{xxx}-\frac{3}{2}v_x^{-5/2}v_{xx}^2
-2\lambda_1v_x^{-1/2}+\frac{2\lambda_2}{3}\right)D_x^{-1}v_x^{-3/2}v_{xx}.
\end{gather}
Under the hodograph transformation (\ref{TheHT}), eq.
(\ref{quasi2_bpot2}) takes the form
\begin{gather}
\label{quasi2_HT2}
V_t=
V_X^{-3/2}V_{XXX}-\frac{3}{2}V_X^{-5/2}V_{XX}^2
-\frac{2\lambda_2}{3} V_X^{-1/2}
+2\lambda_1V_X^{3/2}
-CV_X-\lambda_3.
\end{gather}
We note that with
\begin{gather*}
\lambda_2=3\lambda_1,\qquad C=-\lambda_3
\end{gather*}
eq. (\ref{quasi2_bpot2}) is invariant under the hodograph tansformation
(\ref{TheHT}).
We now apply (\ref{npot}) to (\ref{quasi2_HT2}),
so that eq. (\ref{quasi2_HT2}) becomes
\begin{gather}
W_t=W^{-3/2}W_{XXX}-\frac{9}{2}W^{-5/2}W_XW_{XX}
+\frac{15}{4}W^{-7/2}W_X^3
+\lambda_1 W^{-3/2}W_X\notag\\[0.3cm]
\label{quasi2_W2}
\qquad+\lambda_2W^{1/2}W_X+\lambda_3W_X.
\end{gather}
Equation (\ref{quasi2_W2})
is equivalent to eq. (\ref{gen_quasi2}) of
Proposition 3.1 with
\begin{gather*}
u(x,t)=W^{-1/2}(X,t),\qquad X=x.
\end{gather*}

\strut\hfill

\subsection{A nonlocal recursion operator}

The following equation was introduced in
\cite{Cal-Deg, Ibra-Shab}:
\begin{gather}
\label{CDIS}
\boxed{
u_t=u_{xxx}+3u^2u_{xx}+9uu_x^2+3u^4u_x
}
\end{gather}
which is a special case of our semilinear equation
(\ref{gen_eq}). As pointed out in \cite{Cal87, Sok_Shab},
the linearisation of eq. (\ref{CDIS}) can be achieved
by a nonlocal transformation. We establish this linearisation
using a combination of potential forms and $x$-generalised hodograph
transformations. We then use these transformations to
calculate the recursion operator for (\ref{CDIS}). The recursion
operator for (\ref{CDIS}) turns out to be
nonlocal, which explains why this result is not part of Proposition 2.1.

First we write (\ref{CDIS}) in potential form and set
\begin{gather*}
v_x=u^2.
\end{gather*}
This leads to
\begin{gather}
\label{CDIS-pot}
\boxed{
v_t=v_{xxx}-\frac{3}{4}v_x^{-1}v_{xx}^2+3v_xv_{xx}+v_x^3
}
\end{gather}
Now apply for (\ref{CDIS-pot})
the $x$-generalised hodograph transformation
\begin{gather}
\label{3_hodosemq1}
_3\mbox{\bf H}:\left\{\ba{l}
\displaystyle{dx(X,T)
=
k^{-1}dX}\\[3mm]
\hphantom{dx(\tilde x,\tilde t)}
\displaystyle{
+\left\{-3kk'V_X-\left(\frac{1}{4}(k')^2 +kk''\right)V_X^2
-kk'V_{XX}-k^2\right\}dT}
\\[3mm]
dt(X,T)=dT \\[3mm]
v(x,t)=X,
\ea\right.
\end{gather}
which transforms (\ref{CDIS-pot}) to
\beg
\label{CDIS-xGHT1}
\ba{l}
\displaystyle{
V_T=k^3V_{XXX}+\left(\frac{3}{2}k^2k'+3k^3\frac{k''}{k'}\right)V_XV_{XX}
+3k^3V_{XX}}\\
[0.5cm]
\displaystyle{
\qquad +\left(
\frac{3}{2}k^2k''+k^3\frac{k'''}{k'}\right)V_X^3
+3\left(k^2k'+k^3\frac{k''}{k'}\right)V_X^2+2k^3V_X}.
\ea
\eeq
where $k=k(V)$.
It is remarkable that eq. (\ref{CaseI_bpot}) with
$\lambda_1=\lambda_2=\lambda_3=C=0$, i.e.
\begin{gather}
\label{linearisable}
\tilde V_{\tilde T}=\tilde V_{\tilde X\tilde X\tilde
X}-\frac{3}{4}\tilde V_{\tilde X}^{-1}\tilde V_{\tilde X\tilde X}^2
\end{gather}
can also be transformed to (\ref{CDIS-xGHT1}) by the $x$-generalised
hodograph transformation
\begin{gather}
\label{3_linq1}
_3\mbox{\bf H}:\left\{\ba{l}
\displaystyle{d\tilde X(X,T)
=
k^{-1}dX}\\[3mm]
\hphantom{dx(\tilde x,\tilde t)}
\displaystyle{
+\left\{-3kk'V_X-\left(\frac{1}{4}(k')^2 +kk''\right)V_X^2
-kk'V_{XX}-k^2\right\}dT}
\\[3mm]
d\tilde T(X,T)=dT \\[3mm]
\tilde V(\tilde X,\tilde T)=\frac{1}{2}e^{2X}.
\ea\right.
\end{gather}
Equation (\ref{linearisable}) is the potential form of
the third-order equation
\begin{gather}
\label{lin_W}
W_{\tilde T}=W_{\tilde X\tilde X\tilde X}+3W^{-1}W_{\tilde X}
W_{\tilde X\tilde X},
\end{gather}
which is part of the linearisable hierarchy VIII in \cite{eul_eul_pet}.
The relation of (\ref{lin_W})
to (\ref{CDIS-xGHT1}) is established by
\begin{gather}
W_{\tilde X}=\tilde V^4,
\end{gather}
whereas
the linearisation of (\ref{lin_W})
in \begin{gather}
\label{lin_U}
U_{\tilde T}=U_{\tilde X\tilde X\tilde X}
\end{gather}
is obtained  \cite{eul_eul_pet} by 
\begin{gather}
\label{linearisation_W}
W^2=2U_{\tilde X}.
\end{gather}
The composition of the above transformations
for (\ref{CDIS-pot}) and (\ref{linearisable}), that is
the change of coordinates
\begin{gather*}
(\tilde X,\tilde T,\tilde V(\tilde X,\tilde T))\mapsto
(x,t,v(x,t)),
\end{gather*}
results in the transformation
\begin{gather}
v(x,t)=\frac{1}{2}\ln(2\tilde V),\quad
dx(\tilde X,\tilde T)=d\tilde X,\quad
dt(\tilde X,\tilde T)=d\tilde T.
\end{gather}
Thus the transformation between (\ref{CDIS}) and (\ref{linearisable}),
that is the coordinate transformation
\begin{gather*}
(\tilde X,\tilde T,\tilde V(\tilde X,\tilde T))\mapsto
(x,t,u(x,t)),
\end{gather*}
takes the form
\begin{gather}
\label{lin_final}
u(x,t)=\left(\frac{\tilde V_{\tilde X}}{2\tilde V}\right)^{1/2},
\quad x=\tilde X,\quad t=\tilde T.
\end{gather}
Finally, the linearisation of (\ref{CDIS}) in (\ref{lin_U}) is
given by the following nonlocal transformation:
\begin{gather}
\label{trans_nonloc}
\pde{U(x,t)}{x}=\sqrt{2}\,u\exp\left(\int u^2(x,t)\,dx\right).
\end{gather}
The recursion operator for (\ref{CDIS}) is nonlocal and takes the form
\begin{gather}
R[u]=D_x^2+2u^2D_x+10uu_x+u^4\notag \\
\qquad +2\left(
u_{xx}+2u^2u_x
+2ue^{-2\int u^2dx}\int e^{2\int u^2dx}u_x^2dx\right)D^{-1}u\notag\\
\label{rec_CDIS}
\qquad -2ue^{-2\int u^2dx}D_x^{-1}\left[
\left(u_{xx}+2u^2u_x\right)e^{2\int u^2dx}
+2u\int e^{2\int u^2dx}u_x^2dx\right].
\end{gather}
The recursion operator (\ref{rec_CDIS})
is obtained by transforming the recursion
operator of eq. (\ref{linearisable}), namely
\begin{gather}
\label{rec_tildeV}
R[\tilde V]=\left(
D_{\tilde X}-D_{\tilde X}^{-1}\left[\tilde V^{-1}\tilde V_{\tilde
X}+\tilde V^2\right]D_{\tilde X}\right)^2,
\end{gather}
with the transformations
(\ref{lin_final}).
Note that (\ref{rec_CDIS}) may also be written in
the following nonlinear, but equivalent, form:
\begin{gather}
R[u]=D_x^2+2u^2D_x+10uu_x+u^4+2\left(u_{xx}+2u^2u_x\right)D_x^{-1}u
\notag\\[0.3cm]
\qquad
\label{CDIS_recnonlin}
-2ue^{-2\int u^2dx}D_x^{-1} e^{2\int u^2 dx}
\left[u_{xx}+2u^2u_x-2u_x^2D_x^{-1}u\right].
\end{gather}
To get the next equation in the hierarchy of eq. (\ref{CDIS})
we apply the recursion operator on the $t$-translation symmetry
of  (\ref{CDIS}), i.e. $R[u]u_t$, which leads to the
fifth-order
equation
\begin{gather}
u_t=u_{xxxxx}+5u^2u_{xxxx}+40uu_xu_{xxx}+25uu_{xx}^2+50u_x^2u_{xx}+10u^4u_{xxx}
\notag\\[0.3cm]
\label{CDIS-5}
\qquad
+120u^3u_xu_{xx}+140u^2u_x^3
+10u^6u_{xx}+70 u^5u_x^2+5u^8u_x.
\end{gather}
We recall that the coefficients $I_j$ of the term $I_jD_x^{-1}J_j$
in the linear recursion
operator ({\ref{rec_CDIS}) are symmetries of the given equation. 
Therefore, eq. ({\ref{rec_CDIS}) admits the following
two nonlocal symmetries:
\begin{gather}
Z_1=ue^{-2\int u^2dx}\ \pde{\ }{u}\\
Z_2=\left(
u_{xx}+2u^2u_x
+2ue^{-2\int u^2dx}\int e^{2\int u^2dx}u_x^2dx\right)\pde{\ }{u}
\end{gather}
We can now construct more nonlocal symmetries for (\ref{CDIS})
by applying the
recursion operator on the nonlocal symmetries $Z_1$ and $Z_2$.
For $Z_1$
we get zero, but for $Z_2$ we obtain the fourth order nonlocal symmetry
\begin{gather}
Z_3=\left(u_{xxxx}+4u^2u_{xxx}+28uu_xu_{xx}+8u_x^3+6u^4u_{xx}
+32u^3u_{x}^2+4u^6u_x
\vphantom{-2ue^{-2\int u^2dx}\int e^{2\int u^2dx...}}\right.
\notag\\
\qquad\left.
\label{sym_4th}
-2ue^{-2\int u^2dx}\int e^{2\int u^2dx}\left[u^2_{xx}
+4u^2u_xu_{xx}-4uu_x^3+2u^4u_x^2\right]dx\right)\pde{\ }{u}.
\end{gather}

\section{Conclusion}

We present recursion operators for
a number of integrable equations of the
semilinear form (\ref{gen_eq})
and the quasilinear form (\ref{gen_quasi}). Those are given in
Proposition 2.1 and Proposition 3.1. In addition we report
the potential forms of those equations and apply
a sequence of coordinate transformations, namely the hodograph
transformation and the $x$-generalised hodograph transformation.
In some cases the transformed equations admit local recursion
operators, which are given explicitly. In other cases
the transformed equations do not admit a
recursion operator of the form (\ref{gen_R})
and we suspect that those equations
admit only nonlocal recursion operators and/or recursion operators
which depend explicitly on $x$ and $t$. We do not study 
such recursion operators here.
However, we make one
exception namely for eq. (\ref{CDIS}), for
which we calculate the nonlocal recursion operator by using
coordinate transformations.

Below we give a list of all semilinear and quasilinear
equations (up to $u\to k(u)$) for which we 
obtained recursion operators in this paper:

\strut\hfill

\noindent
{\bf Semilinear equations:}
\begin{gather*}
u_t=u_{xxx}+\lambda_1u_x^3+\lambda_2 u_x^2+\lambda_3u_x+\lambda_4
\end{gather*}
\begin{gather*}
u_t=u_{xxx}+n(u)u_xu_{xx}+m(u)u_x^3+q(u)u_x\\
\qquad q'''-nq''-3n'q'-\frac{2}{3}n^2q'+8mq'=0\\
\qquad 9m'-3n''-6nm+\frac{2}{3}n^3=0
\end{gather*}
\begin{gather*}
u_t=u_{xxx}+3n(u)u_x
u_{xx}-\frac{1}{2}uu_x^3+q(u)u_x\\
\qquad q'''-3nq'' -9\left(n'+\frac{2}{3}n^2+\frac{4}{9}
u\right)q'=0\\
\qquad n''=2n^3+n u-\frac{1}{2}
\end{gather*}
\begin{gather*}
u_t=u_{xxx}+3u^{-1}u_xu_{xx}+ \lambda_1u^4u_x+\lambda_2
u^2u_x+\lambda_3u_x
\end{gather*}
\begin{gather*}
u_t=u_{xxx}-3u^{-1}u_xu_{xx}+\frac{3}{2}u^{-2}u_x^3
+\lambda_1u^2u_x+\lambda_2 u^{-2}u_x+\lambda_3u_x
\end{gather*}
\begin{gather*}
u_t=u_{xxx}-\frac{\beta^2}{8}u_x^3+
\lambda_1e^{\beta u}u_x+\lambda_2 e^{-\beta
u}u_x+\lambda_3u_x
\end{gather*}
\begin{gather*}
u_t=u_{xxx}+\frac{\beta^2}{8}u_x^3+
\lambda_1\cos(\beta u)u_x+\lambda_2 \sin(\beta
u)u_x+\lambda_3u_x
\end{gather*}
\begin{gather*}
u_t=u_{xxx}+\lambda_1 u^2u_x+\lambda_2uu_x+\lambda_3u_x
\end{gather*}
\begin{gather*}
u_t=u_{xxx}+3u^2u_{xx}+9uu_x^2+3u^4u_x
\end{gather*}

\strut\hfill

\noindent
{\bf Quasilinear equations:}
\begin{gather*}
u_t=u^3u_{xxx}+3u^2u_xu_{xx}+\lambda_1u^3u_x+\lambda_2u^2u_x+\lambda_3u_x
\end{gather*}
\begin{gather*}
u_t=u^3u_{xxx}+\lambda_1u^3u_x+\lambda_2u^{-1}u_x+\lambda_3u_x
\end{gather*}

\strut\hfill

\noindent
{\bf Linearisable third-order equations:}\n

\noindent
For the reader's convenience
we also list here the linearisable third-order evolution
equations which admit $x$- and $t$-independent recursion operators given in
\cite{eul_eul_pet}.
Below $\beta,\ \gamma,\ \alpha$ are arbitrary constants
and $k$ a nonconstant ${\cal C}^3$ function of $u$:

\begin{gather*}
u_t=u_{xxx}+3\left(\frac{k''}{k'}+\alpha k'\right)u_xu_{xx}
+\left(\frac{k'''}{k'}+3\alpha k''+\alpha^2(k')^2\right)u_x^3\\[0.3cm]
\qquad+\alpha\gamma u_x+\beta\left(u_{xx}+
\left(\frac{k''}{k'}+\alpha
k'\right)u_x^2\right).
\end{gather*}
\begin{gather*}
u_t=u_{xxx}+3\frac{k''}{k'}u_xu_{xx}+\frac{k'''}{k'}u_x^3
+\frac{3\alpha}{2}ku_{xx}
+\frac{3\alpha}{2}\left(k\frac{k''}{k'}+k'\right)u_x^2\\[0.3cm]
\qquad +\frac{3\alpha^2}{4}k^2u_x
+\beta\left(u_{xx}+\frac{k''}{k'}u_x^2+\alpha ku_x\right)
\end{gather*}
\begin{gather*}
u_t=k^3u_{xxx}+3\left(k^3\frac{k''}{k'}-k^2k'\right)u_xu_{xx}
+\left(k^3\frac{k'''}{k'}+3k^2k''\right)u_x^3
+3\alpha k^3u_{xx}\\[0.3cm]
\qquad
+3\alpha\left(k^3\frac{k''}{k'}+k^2k'\right)u_x^2+2\alpha^2k^3u_x.
\end{gather*}
The above three equations are the second equations in the
linearisable hierarchies
(\ref{Eq2_1}), (\ref{Eq2_2}) and (\ref{Eq2_3}), respectively.
Their recursion operators are
(\ref{R2_1}), (\ref{R2_2}) and (\ref{R2_3}), respectively.

\label{euler-lastpage}

\end{document}